# Polyphonic audio tagging with sequentially labelled data using CRNN with learnable gated linear units

*Yuanbo Hou*[1], *Qiuqiang Kong*[2], *Jun Wang*[1], *Shengchen Li*[1]

[1] Beijing University of Posts and Telecommunications, Beijing, P.R.China
{hyb, shengchen.li, wangjun19930314}@bupt.edu.cn
[2] Centre for Vision, Speech and Signal Processing, University of Surrey, UK
q.kong@surrey.ac.uk

**ABSTRACT**

Audio tagging aims to detect the types of sound events occurring in an audio recording. To tag the polyphonic audio recordings, we propose to use Connectionist Temporal Classification (CTC) loss function on the top of Convolutional Recurrent Neural Network (CRNN) with learnable Gated Linear Units (GLU-CTC), based on a new type of audio label data: Sequentially Labelled Data (SLD). In GLU-CTC, CTC objective function maps the frame-level probability of labels to clip-level probability of labels. To compare the mapping ability of GLU-CTC for sound events, we train a CRNN with GLU based on Global Max Pooling (GLU-GMP) and a CRNN with GLU based on Global Average Pooling (GLU-GAP). And we also compare the proposed GLU-CTC system with the baseline system, which is a CRNN trained using CTC loss function without GLU. The experiments show that the GLU-CTC achieves an Area Under Curve (AUC) score of 0.882 in audio tagging, outperforming the GLU-GMP of 0.803, GLU-GAP of 0.766 and baseline system of 0.837. That means based on the same CRNN model with GLU, the performance of CTC mapping is better than the GMP and GAP mapping. Given both based on the CTC mapping, the CRNN with GLU outperforms the CRNN without GLU.

## 1. INTRODUCTION

Audio tagging aims to detect the types of sound events occurring in an audio recording. Audio recordings are typically short segments such as the audio recordings in IEEE AASP DCASE 2018 Challenge Task 4 [1]. Audio tagging has many applications in information retrieval [2], audio classification [3], acoustic scene recognition [4] and industry sound recognition [5].

Most previous works of audio tagging relies on strongly labelled data or weakly labelled data. In strongly labelled data [4], each audio clip is labelled with both the tags and the onset and offset times of sound events. However, labelling strong label is time consuming and labor expensive, resulting strongly labelled data is scarce and its size is often limited to minutes or a few hours [6]. Thus the audio research community have turned to large-scale datasets without the onset and offset times of sound events, which is referred to as Weakly Labelled Data (WLD) [7]. WLD is also called clip level labelled data. In WLD, only the presence or absence of sound events are known, but the occurrence sequence of sound events are not known.

In this paper, we explore the possibility of Sequentially Labelled Data (SLD) in real-life polyphonic audio tagging. SLD is a type of audio label newly proposed in [8]. In SLD, both the tags of audio clip and the sequence of tags are known, without the onset and the offset of tags. SLD reduces the workload of data annotation and avoids the problem of inaccurate onset and offset annotation of tags in strongly labelled data. In addition, SLD contains the sequential information of tags which is not provided by WLD [8]. However, in the previous work [8], the SLD was the synthesized monophonic audio based on IEEE DCASE 2013 development dataset, there is no overlap between sound events. To explore the possibility of SLD in real-life audio recordings, we manually label 1578 polyphonic audios of DCASE 2018 Task 4 with sequential labels and release it here[1]. The details of sequential labelling of polyphonic audio recordings will be introduced in Section 3.

To predict the sequential labels of SLD in polyphonic audio recordings, we propose to use CTC loss function on the top of CRNN with learnable Gated Linear Units (GLU-CTC). This idea is inspired by the great performance of CTC in Automatic Speech Recognition [9]. CTC is a learning technique for sequence labelling with RNN, which allows RNN to be trained for sequence-to-sequence tasks without requiring any prior alignment between the input and target sequences. In GLU-CTC, CTC objective function maps the frame-level probability of sound events to the target sequential labels of sound events, similar to the pooling layer in neural networks. So we explore the performance of this three pooling function: CTC, Global Max Pooling (GMP) and Global Average Pooling (GAP) in polyphonic audio tagging, based on the same CRNN with GLU. This three models are abbreviated as GLU-CTC, GLU-GMP and GLU-GAP, respectively. In this paper, the baseline system is a CRNN without GLU train with CTC loss function.

There are two contributions in this paper. First, in polyphonic audio tagging we explore the possibility of a new label type: Sequentially Labelled Data, which not only reduces the workload of data annotation in strong labels, but also indicates the sequential information of tags in weak labels. We release the SLD of DCASE 2018 Task 4 in here[1]. Second, to predict the sequential labels of SLD in polyphonic audio recordings, we propose to use CTC learning technique to train a CRNN model with learnable GLU. And we compare the performance of GLU-CTC, GLU-GMP, GLU-GAP and the baseline system, which is a CRNN train with CTC loss function. There is no GLU in baseline system.

---

[1] https://github.com/moses1994/DCASE2018-Task4

This paper is organized as follows, Section 2 introduces related works. Section 3 describes the annotation method of SLD in polyphonic audio recordings. Section 4 describes how the CTC uses SLD in polyphonic audio tagging and the model structure. Section 5 describes the dataset, experimental setup and results. Section 6 gives conclusions.

## 2. RELATED WORK

Audio classification and detection have obtained increasing attention in recent years. There are many challenges for audio detection and tagging such as IEEE AASP challenge on DCASE 2013 [4], DCASE 2016 [10] and DCASE 2017 [6].

Many conventional works of audio classification and audio clip tagging used Mel Frequency Cepstrum Coefficient (MFCC) and Gaussian Mixture Model (GMM) as baseline system [4]. Recent audio classification methods including Deep Neural Networks (DNNs) [6], Convolution Neural Networks (CNNs) [11] and Recurrent Neural Networks (RNN) [3], with inputs varying from Short-Time Fourier Transform (STFT), Mel energy, spectrogram, MFCC to Constant Q Transform (CQT) [12].

The bag of frames (BOF) model was used in [13], where an audio clip is cut into segments and each segment inherits the labels of the audio clip. BOF is based on an assumption that tags occur in all frames, which is however not the case in practice. Some sound events such as "gunshot" only happen a short time in an audio clip. State-of-the-art audio tagging methods [14] transform waveform to the Time-Frequency (T-F) representation. Then, the T-F representation is treated as an image which is fed into CNNs. However, unlike image where the objects usually occupy a dominant part of an image, in an audio clip events only occur a short time. To solve this problem, attention models [15] for audio tagging and classification are applied to attend to the audio events and ignore the back ground sounds.

## 3. SEQUENTIALLY LABELLED DATA

The polyphonic audio data used in this paper is the weak annotations training set of DCASE 2018 Task 4, a subset of Google Audio Set [16]. Audio Set consists of an ontology of 632 sound event classes and a collection of 2 million human-labeled 10-second audio clips drawn from YouTube [16].

In the training set, the polyphony makes it hard to define ordered sequences of sound events. To tackle this problem, we use the order of boundaries of each sound event, the order of onset and offset, but not the time stamps as the sequential labels. For example, we could use the sequential labels *dishes_start, dishes_end, dishes_start, dishes_end, speech_start, blender_start, speech_end, speech_start, blend_end, speech_end* as the sequential label for the audio clip in Fig. 1. Another example is if the content of an audio clip could be described by *a dog barks while a car rings*, we used the sequential labels *ring_start, dog_start, dog_end, ring_end* as the sequential label. In the ground truth label sequence, the tags of the audio clip and the sequence of tags are known, without knowing their occurrence time. We refer to the audio clip labelled by label sequence as Sequentially Labelled Data (SLD). Fig. 1 shows an audio clip with strong, sequential and weak tags.

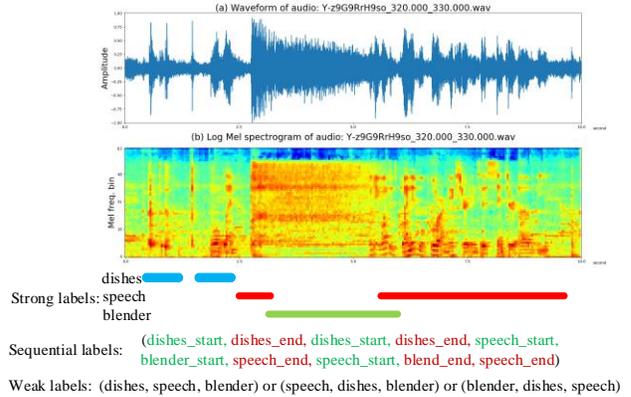

Figure 1: From top to bottom: (a) waveform of an audio clip containing three sound events: "dishes, speech, blender"; (b) log Mel spectrogram of (a); Strong labels, sequential labels and weak labels of the audio clip.

In this paper, we manually labelled the weak annotations training set of DCASE 2018 Task 4 with sequential labels and release it after verification. See here[1] for more details about SLD.

## 4. METHOD

In this section, we will explain how to use CTC in polyphonic audio tagging based on SLD. Then, we will describe the model structure used in this paper.

### 4.1. CTC in Polyphonic Audio Tagging using SLD

CTC is a learning technique for sequence labelling, it shows a new way for training RNN with unsegment sequences. In fact, CTC redefines the loss function of RNN [17] and allows RNN to be trained for sequence-to-sequence tasks, without requiring any prior alignment (*i.e.* starting and ending time of sound events) between the input and target sequences [9]. In audio tagging, we are only interested in the label sequence of corresponding audio clip, not the ground truth alignment of events in the audio clip. Thus, we want to marginalize out the alignment.

To marginalize out the alignment, first, CTC adds an extra "blank" label (denoted by "-") to original label set $L$ [9]. Then, it defines a many-to-one mapping $\beta$ that transforms the alignment (*i.e.* the sequence of output labels at each time step, also called a path [17]) to label sequence. The mapping β removes repeated labels from the path to a single one, then removes the "blank" labels. For example, β(C-AT-)=β(-CC--ATT)=CAT, that is, path 'C-AT-' and '-CC--ATT' both map to the label sequence 'CAT'.

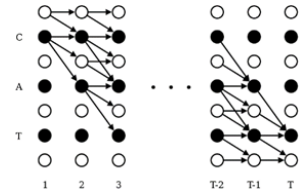

Figure 2: Trellis for computing CTC loss function [17] applied to labelling 'CAT'. Black circles represent labels, white circles represent blanks. Arrows signify allowed transitions.

The CTC objective function is defined as the negative logarithm of the total probability of all paths [9] that map to the ground truth label sequence. The total probability can be found using dynamic programming algorithm [17] on the trellis shown in Fig. 2. On the x-axis is time steps, on the y-axis is "modified label sequence", that is target label sequence with blank labels added to the beginning and the end and inserted between every pair of labels.

When we use the simple best path decoding to decode the output of CTC, the output of CTC is directly the label sequence. By this means no threshold is needed to determine whether there are corresponding events in the audio clip. This will reduce the risk of over-fitting due to specific thresholds, which is an advantage of using CTC loss function in audio tagging. More details about CTC can be seen [17].

### 4.2. Model Structure

Inspired by the good performance of CRNN in audio tagging [15], CRNN is used in this paper shown in Fig. 3. First, the waveforms of audio clips are transformed to T-F representations such as Mel spectrograms. And convolutional layers are applied on the T-F representations to extract high level features. Next, Bidirectional Gated Recurrent Units (BGRU) are adopted to capture the temporal context information. Finally, the output layer is a dense layer with sigmoid activation function since audio tagging is a multi-class classification problem [3, 6].

In the CRNN, the output activation from the CNN layers are padded with zeros to keep the dimension of the output the same as input. And the max-pooling is applied in the frequency axis only to preserve the time resolution of the input. Clip level probability of tags can be obtained from the last layer. To compare the performance of different pooling function, there are three pooling operations in Fig. 3, CTC, Global Max Pooling (GMP) and Global Average Pooling (GAP).

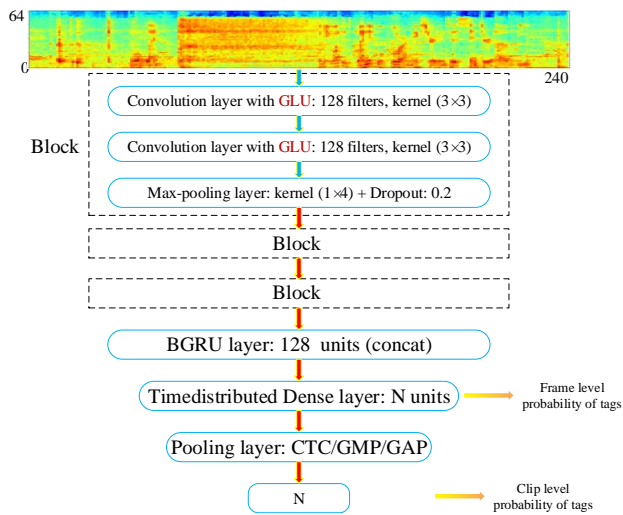

Figure 3: Model Structure. Due to the acoustic event classes number is 10 in DCASE 2018 Task 4, thus, for model with GMP and GAP layer, N=10. For model with CTC layer, N=21 (10 *2+1), the extra '1' indicates the blank label.

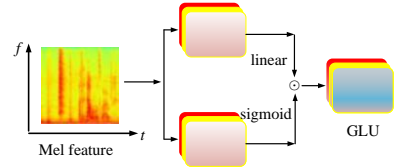

Figure 4: The Structure of GLU.

### 4.3. Gated Linear Units

As shown in Fig. 3, a CRNN model with 13 layers is applied for audio tagging. In order to reduce the gradient vanishing problem in deep networks, the Gated Linear Units (GLU) [18] is used as the activation function to replace the ReLU [19] activation function in the CRNN model. The structure of GLU is shown in Fig. 4. By providing a linear path for the gradients propagation while keeping nonlinear capabilities through the sigmoid operation, GLU can reduce the gradient vanishing problem for deep networks [18]. Similar to the gating mechanisms in long short-term memories [20] or gated recurrent units [21], GLU can control the amount of information of a T-F unit flow to the next layer. GLU are defined as:

$$Y = (W * X + b) \odot \sigma(V * X + c) \quad (1)$$

where $\sigma$ is sigmoid function, the symbol $\odot$ is the element-wise product and $*$ is the convolution operator. $W$ and $V$ are convolutional filters, $b$ and $c$ are biases. $X$ denotes the input T-F representation in the first layer or the feature maps of the interval layers in model.

The value of sigmoid function ranges from 0 to 1, so if a GLU gate value is close to 1, then the corresponding T-F unit is attended. If a GLU gate value is near to 0, then the corresponding T-F unit is ignored. By this means the network can learn to attend to sound events and ignore the unrelated sounds.

## 5. EXPERIMENTS AND RESULTS

### 5.1. Dataset, Experiments Setup and Evaluation Metrics

In this paper, the training set is 1578 clips (2244 class occurrences) of Task 4 from domestic environments, which consists of 10 classes of sound events. We manually label the 1578 audio clips with sequential labels and release it after verification, the annotation method is described in Section 3. The test set is 288 polyphonic audio clips (906 events) of Task 4 [1].

For all the models described in this paper, in training, log Mel band energy is extracted in Hamming window of length 64 ms with 64 Mel frequency bins [22]. For a given audio clip of 10-second in Task 4, this feature extraction block results in a (240×64) output as shown in Fig. 3. 240 is the number of frames and 64 is the number of Mel frequency bins. The binary cross-entropy loss [23] is applied between the predicted probability of each tag and the corresponding ground truth tag. Dropout and early stopping criteria are used in training phase to prevent overfitting. The model is trained for maximum 200 epochs with Adam optimizer with a learning rate of 0.001.

Table 1: Averaged Stats of Audio Tagging

| Metric | AUC of each event class | | | | | | | | | | avg. | | | |
|---|---|---|---|---|---|---|---|---|---|---|---|---|---|---|
| Event | Speech | Dog | Cat | Bell | Dishes | Frying | Blender | Water | cleaner | Shaver | AUC | Precision | Recall | F-score |
| GLU-GAP | 0.895 | 0.946 | 0.875 | 0.820 | 0.583 | 0.602 | 0.641 | 0.773 | 0.771 | 0.758 | 0.766 | **0.960** | 0.588 | 0.730 |
| GLU-GMP | 0.909 | 0.946 | 0.921 | 0.873 | 0.669 | 0.643 | 0.691 | 0.813 | 0.785 | 0.778 | 0.803 | 0.957 | 0.645 | 0.771 |
| GLU-CTC | **0.941** | **0.969** | **0.994** | **0.942** | **0.762** | **0.905** | 0.753 | **0.860** | **0.850** | **0.835** | **0.882** | 0.816 | **0.816** | **0.816** |
| Baseline | 0.912 | 0.953 | 0.957 | 0.836 | 0.684 | 0.776 | **0.795** | 0.839 | 0.808 | 0.808 | 0.837 | 0.706 | 0.763 | 0.734 |

To evaluate the results of audio tagging in clip level in this paper, we follow the metrics proposed in [22]. The results are evaluated by *Precision* ($P$), *Recall* ($R$) and *F-score* [24] and Area Under Curve (*AUC*) [25]. Larger $P$, $R$, *F-score* and *AUC* indicates better performance.

### 5.2. Results

In this paper, the GLU-CTC, GLU-GMP and GLU-GAP all contain the learnable GLU, which introduces the attention mechanism in the convolutional layers in CRNN. However, there is no GLU in the baseline model, which is a CRNN train with CTC objective function. To evaluate the performance of the models in this paper, we calculate the *AUC* score of audio tagging results in clip level of these models. As shown in Table 1, GLU-CTC achieves an averaged *AUC* of 0.882 outperforming the GLU-GAP and GLU-GMP, and also better than the baseline system. Table 1 also shows the averaged statistic including *Precision*, *Recall*, *F-score* and *AUC* over 10 kinds of sound events, respectively. GLU-CTC mapping performs better than GLU-GAP and GLU-GMP, too. That is, based on the same CRNN model with GLU, the performance of CTC mapping function is better than the GAP and GMP mapping function in polyphonic audio tagging.

The averaged stats of audio tagging is evaluated in clip level of audio clips, the frame level predictions of models on example audio clip was shown in Fig. 5. In Fig. 5, the predictions of GLU-GAP in frame level is always 1, which means the predictions of GLU-GAP in frame level overestimates the occurrence probability of corresponding event. While GLU-GMP, in contrast, underestimates it. GLU-GMP produces wide peaks, indicating the onset and offset times of event. That shows max pooling has ability to locate event, while average pooling seems to fail. The reason may be max pooling encourages the response for a single location to be high [26], for similar audio events which can obtain similar features. While average pooling encourages all response to be high [26], difference features of each event make it difficult to locate event.

In Fig. 5, the GLU-CTC could predict the onset (start) and offset (end) tag sequence of corresponding audio recording, typically as a series of spikes [17]. Although the spikes align well with the actual position of the boundaries of sound events in audio recording, there is no time span information about these events. The spikes outputted by GLU-CTC could locate corresponding events in the audio clip, while baseline system seems to fail, which means the attention mechanism introduced by GLU is helpful for audio tagging. The reason may be the attention introduced by GLU focuses on the local information within each feature map, which could help GLU-CTC better learn the high-level representations of corresponding audio events.

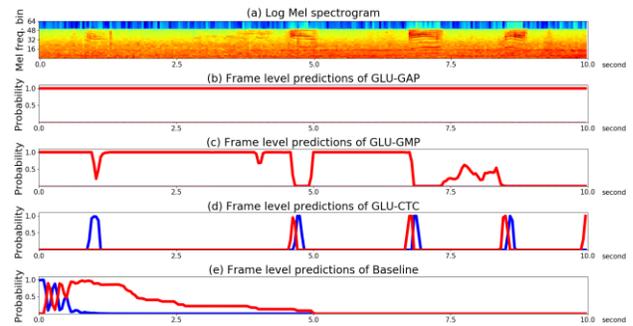

Figure 5: Frame level predictions of GLU-GAP (b), GLU-GMP (c), GLU-CTC (d), and Baseline (e). In GLU-CTC and Baseline, blue peaks denote the starting and red peaks denote the ending of corresponding sound events.

### 6. CONCLUSION

In this paper, we explore the possibility of a new type of audio label data called SLD in polyphonic audio tagging. To utilize SLD in audio tagging, we propose a GLU-CTC model. In GLU-CTC, CTC layer maps frame level tags to clip level tags, similar to the pooling operations. Experiments show GLU-CTC outperforms GLU-GAP and GLU-GMP. Finally, we released the sequential labels of DCASE 2018 Task 4 after verification. In the future, we will explore the possibility of SLD in sound event detection with polyphonic audio recordings and try to expand the size of SLD.

### 7. ACKNOWLEDGMENT

Qiuqiang Kong was supported by EPSRC grant EP/N014111/1 ``Making Sense of Sounds" and a Research Scholarship from the China Scholarship Council (CSC) No. 201406150082."

# 8. REFERENCES

[1] http://dcase.community/challenge2018/task-large-scale-weakly-labeled-semi-supervised-sound-event-detection.

[2] G. Guo and Stan Z Li, "Content-based audio classification and retrieval by support vector machines," *IEEE Transactions on Neural Networks*, vol. 14, no. 1, pp. 209–215, 2003

[3] Y. Xu, Q. Kong and W. Wang, et al. "Large-scale weakly supervised audio classification using gated convolutional neural network," arXiv preprint arXiv: 1710.00343, 2017.

[4] D. Stowell, D. Giannoulis and E. Benetos, et al. "Detection and classification of acoustic scenes and events," *IEEE Transactions on Multimedia*, vol. 17, no. 10, pp. 1733–1746, 2015

[5] S. Dimitrov, J. Britz, B. Brandherm, and J. Frey, "Analyzing sounds of home environment for device recognition," in *AmI. Springer*, 2014, pp. 1–16.

[6] A. Mesaros, T. Heittola, A. Diment and B. Elizalde, et al. "DCASE 2017 challenge setup: Tasks, datasets and baseline system," in *Proceedings of DCASE2017 Workshop*.

[7] A. Kumar and B. Raj, "Audio event detection using weakly labelled data," in *Proceedings of the 2016 ACM on Multimedia Conference. ACM*, 2016, pp. 1038–1047

[8] https://www.researchgate.net/publication/326588286_Audio_Tagging_With_Connectionist_Temporal_Classification_Model_Using_Sequentially_Labelled_Data

[9] A. Graves and N. Jaitly, "Towards end-to-end speech recognition with recurrent neural networks", in *Proc. of ICML*, 2014.

[10] M. Valenti, A. Diment and G. Parascandolo, et al., "DCASE 2016 acoustic scene classification using convolutional neural networks," *Workshop on Detection and Classification of Acoustic Scenes and Events (DCASE 2016)*, Budapest, Hungary, 2016

[11] Y. Han and K. Lee, "Acoustic scene classification using convolutional neural network and multiple-width frequency-delta data augmentation," arXiv preprint arXiv: 1607.02383, 2016.

[12] T. Lidy and A. Schindler, "CQT-based convolutional neural networks for audio scene classification," in *Workshop on Detection and Classification of Acoustic Scenes and Events (DCASE 2016)*, Budapest, Hungary, 2016

[13] J. Ye, T. Kobayashi, M. Murakawa, and T. Higuchi, "Acoustic scene classification based on sound textures and events," in *Proceedings of ACM on Multimedia Conference. ACM*, 2015, pp. 1291–1294.

[14] K. Choi, G. Fazekas, and M. Sandler, "Automatic tagging using deep convolutional neural networks," arXiv preprint arXiv: 1606.00298, 2016.

[15] Y. Xu, Q. Kong, Q. Huang, W. Wang, and M. D. Plumbley, "Attention and localization based on a deep convolutional recurrent model for weakly supervised audio tagging," in *INTERSPEECH. IEEE*, 2017, pp. 3083–3087.

[16] Gemmeke, Jort F., et al. "Audio Set: An ontology and human-labeled dataset for audio events." *IEEE International Conference on Acoustics, Speech and Signal Processing IEEE*, 2017:776-780.

[17] Graves A, Gomez F. Connectionist temporal classification: labelling unsegmented sequence data with recurrent neural networks[C]. *International Conference on Machine Learning. ACM*, 2006:369-376.

[18] Y. N. Dauphin, A. Fan, M. Auli, and D. Grangier, "Language modelling with gated convolutional networks," arXiv preprint arXiv: 1612.08083, 2016.

[19] V. Nair and G. E. Hinton, "Rectified linear units improve restricted boltzmann machines," in ICML, 2010, pp. 807–814.

[20] S. Hochreiter and J. Schmidhuber, "Long short-term memory," Neural Computation, vol. 9, no. 8, pp. 1735–1780, 1997.

[21] J. Chung, C. Gulcehre, K. Cho, and Y. Bengio, "Empirical evaluation of gated recurrent neural networks on sequence modeling," arXiv preprint arXiv: 1412.3555, 2014.

[22] Kong Q, Xu Y, Sobieraj I, et al. Sound Event Detection and Time-Frequency Segmentation from Weakly Labelled Data, arXiv preprint arXiv: 1804.04715, 2018.

[23] F. Ghazani, and M. S. Baghshah. "Multi-label classification with feature-aware implicit encoding and generalized cross-entropy loss." *Electrical Engineering IEEE*, 2016:1574-1579.

[24] A. Mesaros, T. Heittola, and T. Virtanen, "Metrics for polyphonic sound event detection," *Applied Sciences*, vol. 6, no. 6, p. 162, 2016.

[25] J. A. Hanley and B. J. McNeil, "The meaning and use of the area under a receiver operating characteristic (roc) curve." *Radiology*, vol. 143, no. 1, pp. 29–36, 1982.

[26] Kolesnikov, Alexander, and C. H. Lampert. "Seed, Expand and Constrain: Three Principles for Weakly-Supervised Image Segmentation." *European Conference on Computer Vision Springer International Publishing*, 2016:695-711.